\documentclass[prl, amsmath,amssymb, amsfonts, preprint]{revtex4-1}
\pdfoutput=1
\usepackage{graphicx}
\usepackage{epstopdf}
\usepackage[margin=0.75in]{geometry}

\begin{document}
  \title{Label-free optical detection of single enzyme-reactant reactions and associated conformational changes}

\author{Eugene Kim}
\author{Martin D. Baaske}
\author{Isabel Schuldes}
\author{Peter S. Wilsch}
\author{Frank Vollmer*}

\affiliation{Max Planck Institute for the Science of Light,
Staudtstra{\ss}e 2, 91058 Erlangen, Germany}

\begin{abstract}
Monitoring the kinetics and conformational dynamics of single enzymes is crucial in order to better understand their biological functions as these motions and structural dynamics are usually unsynchronized among the molecules. Detecting the enzyme-reactant interactions and associated conformational changes of the enzyme on a single molecule basis, however, remain as a challenge with established optical techniques due to the commonly required labeling of the reactants or the enzyme itself. The labeling process is usually non-trivial and the labels themselves might skew the physical properties of the enzyme. Here we demonstrate an optical, label-free method capable of observing enzymatic interactions and the associated conformational changes on the single molecule level. We monitor polymerase/DNA interactions via the strong near-field enhancement provided by plasmonic nanorods resonantly coupled to whispering gallery modes in microcavities. Specifically, we employ two different recognition schemes: one in which the kinetics of polymerase/DNA interactions are probed in the vicinity of DNA-functionalized nanorods, and the other in which these interactions are probed via the magnitude of conformational changes in the polymerase molecules immobilized on nanorods. In both approaches we find that low and high polymerase activities can be clearly discerned via their characteristic signal amplitude and signal length distributions. Furthermore, the thermodynamic study of the monitored interactions suggests the occurrence of DNA polymerization. This work constitutes a proof-of-concept study of enzymatic activities via plasmonically enhanced microcavities and establishes an alternative and label-free method capable of investigating structural changes in single molecules.
\end{abstract}
\maketitle
Enzymes fulfill a plethora of metabolic functions in all living organisms. In many cases, enzymatic activity is closely connected to changes in the enzymes' conformation often involving the transition through multiple substates. One of the most important and perhaps most studied enzymes is DNA polymerase present in all cells and responsible for replicating genetic information. Outside of the actual metabolisms, it is utilized for important biological applications such as the polymerase chain reaction (PCR) and DNA sequencing. The enzymatic activity of DNA polymerase involves multiple steps, such as the binding of primer-hybridized template DNA, insertion of a deoxynucleoside triphosphate (dNTP), and incorporation of dNTP, thereby, extending the strand by one nucleotide. Each step of such a catalytic process is accompanied by conformational changes of the DNA polymerase. These transitions, together with the corresponding reaction pathways, have an intrinsically transient nature and have been vastly studied via single-molecule based techniques. The perhaps most widely used method is single-molecule F\"orster Resonance Energy Transfer (smFRET), which resolves the dynamics of DNA/polymerase interactions and the associated structural changes via measuring distances between labels attached to specific polymerase domains or DNA strands \cite{farooq2014studying,christian2009single,santoso2010conformational,hohlbein2013conformational,berezhna2012single}. Despite its great contribution to the extension of knowledge on the reaction mechanisms of DNA polymerase, this method intrinsically requires chemical modification of the enzyme in order to attach labels, hence, it can cause the studied enzyme to deviate from its natural kinetics. Inherent physical processes such as photobleaching and large background signals also limit the applicability of FRET \cite{rees2005pharmaceutical}. 

In this study we establish a label-free optical sensor platform capable of monitoring the kinetics and conformational dynamics of single enzymes. We show that the kinetics of single-molecule DNA/polymerase (sm-DNA/Pol) interactions as well as the related conformational transitions of polymerase can be studied by using plasmonically enhanced whispering gallery mode (WGM) microcavity sensors \cite{santiago2011nanoparticle, shopova2011plasmonic,swaim2011detection,foreman2013theory,dantham2013label, baaske2014single, baaske2016, kim2016}. Our sensor recognizes conformational changes and the motion of single enzymes as shifts of the cavity's optical resonance wavelength induced by the perturbation of the highly localized electric field at the tips of nanorods (NRs). 
The magnitude and sign of these shifts are proportional to the change in the electric field intensity integrated over the volume of the molecule. For an increasing (decreasing) integrated intensity the induced resonance shift is towards longer (shorter) wavelength. This can be applied for the study of molecular kinetics as following. When a molecule enters the enhanced electric near field in the vicinity of the NRs, it causes a spectral red shift with an increasing magnitude as it moves towards the field's intensity maximum. This is followed by a blue shift with the same magnitude when the molecule moves away from the NRs and leaves the near field completely. This concept also holds true when a molecule immobilized on the NRs changes its shape (i. e. conformational state) in a manner that the change in the volume integrated field intensity is sufficient to cause recognizable spectral shifts of the WGM's position in either direction.  
Here we utilize these mechanisms to probe sm-DNA/Pol interactions with two different approaches: immobilization of DNA on the NRs for the study of sm-DNA/Pol interaction kinetics and immobilization of polymerase on the NRs for the observation of specific conformational transitions accompanied by its interaction with DNA molecules.
In both cases, the statistical analysis of our sensor's signals allows us to discern the different activity levels of three polymerase species: the Klenow fragment of \textit{E. coli} DNA polymerase I (KF) and DNA polymerases from \textit{Thermus aquaticus} (Taq) and \textit{Pyrococcus furiosus} (Pfu). We also study the sm-DNA/Pol interaction kinetics and the associated conformational changes with respect to the type of DNA (primer/template and single-stranded DNA), temperature, and the presence of dNTPs.

\subsection*{Method for monitoring single molecule DNA/polymerase interactions}
\begin{figure*}[htbp]	
\begin{center}
\includegraphics{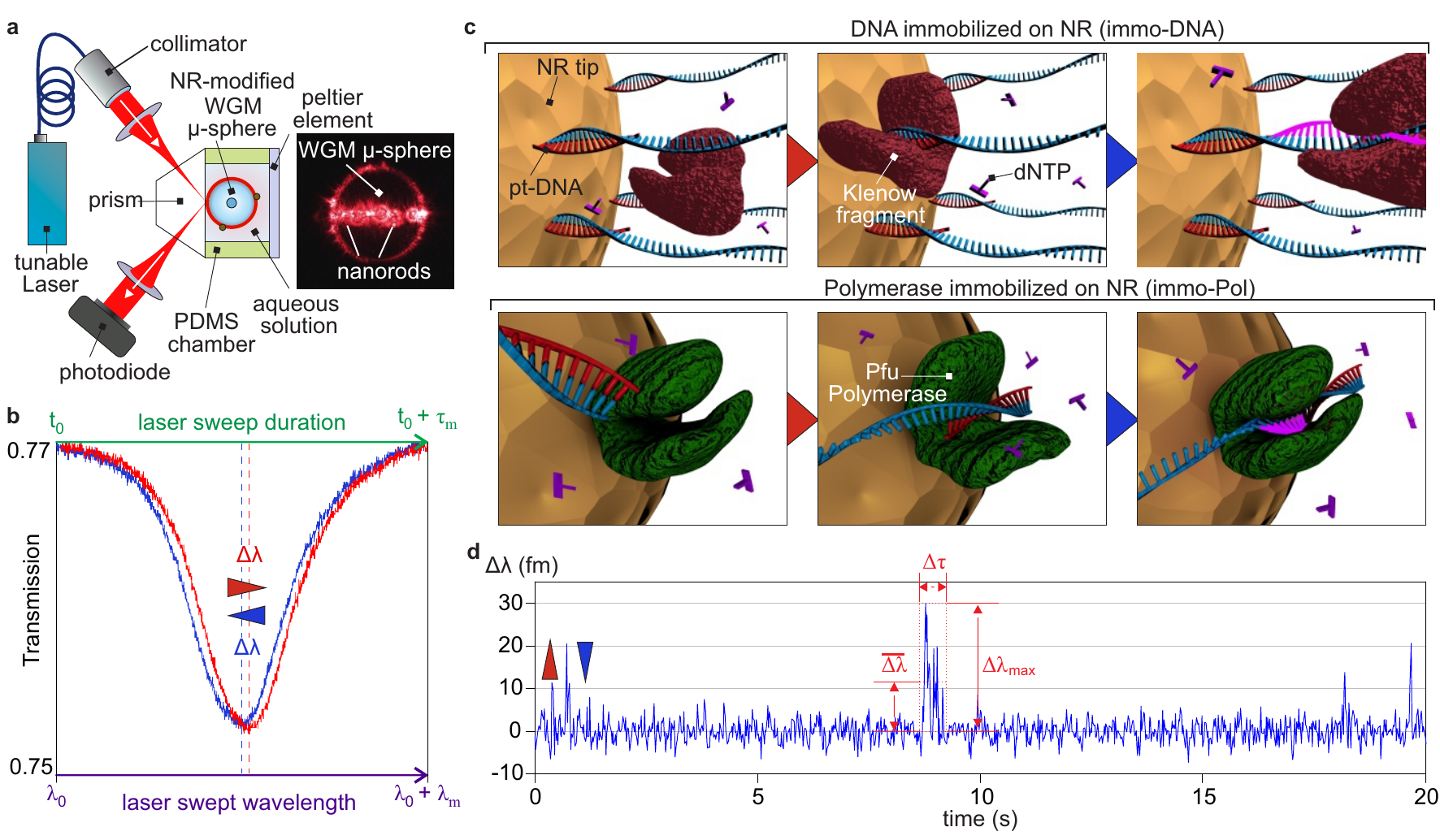}
	\caption{\textbf{Methods for detection of sm-DNA/Pol interactions.} (a) Schematic of the prism-based microcavity sensor setup. The inset shows an image of NR scatterers bound onto the equatorial plane of a microsphere. (b) Typical transmission spectra showing a WGM (Lorentzian dip) before (blue) and during (red) a DNA/polymerase interaction. (c) Conceptual representation of the two different approaches employed for monitoring DNA/polymerase interactions (immo-DNA and immo-Pol scheme) and (d) the corresponding resonance traces, exhibiting spike signals caused by the respective DNA/polymerase interactions.}
\end{center}
\end{figure*} 
The experimental setup used for monitoring interaction kinetics between polymerase and DNA is depicted in Fig. 1a. A fused silica microsphere with a diameter of $\sim$ 80 - 100 $\mu$m serves as a WGM resonator and is placed inside a liquid sample cell made of polydimethylsiloxane (PDMS). A peltier element, which is attached to the wall of the sample cell, allows for temperature regulation of the liquid.  WGMs are excited via frustrated total internal reflection of a wavelength-tunable laser beam ($\lambda_{c}$ $\sim 642$, $780\,\mathrm{nm}$) focused onto the surface of a prism. The resonance wavelengths of WGMs are then determined from transmission spectra obtained by sweeping the laser wavelength with a frequency of 50 Hz via a modified centroid method \cite{baaske2016}. 
Single molecule sensitivity can then be achieved by utilizing the plasmonic near field enhancement provided by gold nanorods, whose longitudinal surface plasmon resonances match the laser's wavelength \cite{zijlstra2012optical, ament2012single, dantham2013label, baaske2014single, baaske2016, kim2016}. 
For this, the NRs are immobilized onto the resonator (See Methods section for the chemical protocols) in a directly monitored process, allowing not only to count the number of deposited NRs, but also to determine if their long axes are aligned reasonably parallel to the polarization of electrical field, via the binding-induced linewidth broadening and wavelength shifts of the monitored WGM (Section 1 of Supplementary Information). The local perturbations of the electric field near the NRs, as induced by sm-DNA/Pol interactions, can then be recognized as shifts $\Delta\lambda$ in the spectral position of WGMs (Fig. 1b). As afore mentioned the magnitude and sign of these shifts are proportional to the changes in the electric field intensity integrated over the volume occupied by the molecule $V_m(t)$ at the times $t_1$ and $t_2=t_1+\Delta t$ (where $\Delta t\approx 20$\,ms is the time between two laser sweeps), as well as to the molecule's polarizability in excess to the medium $\alpha_e$ (assuming a constant and isotropic molecular polarizability) \cite{arnold2003shift, swaim2011detection, shopova2011plasmonic}: 
\begin{equation}
\Delta\lambda\propto\alpha_e \left(\int_{V_m(t_2)} \vert E(r) \vert ^2 \mathrm{d}V-\int_{V_m(t_1)} \vert E(r) \vert ^2 \mathrm{d}V \right)=\alpha_e (I(t_2)-I(t_1))=\alpha_e \Delta I.
\end{equation} 
This, however, does not consider that the process of sweeping the laser over the spectral range occupied by the resonant mode (indicated as $\lambda_m$ in Fig. 1b) itself requires a certain time $\tau_m$. 
In consequence each experimentally measured integrated intensity $I_{exp,k}$ for the k-th sweep of the laser originates from an averaging process:
\begin{equation} 
I_{exp,k}=\overline{I}_k = (\tau_m)^{-1}\int_{t_0}^{t_0+\tau_m}I(t)\mathrm{d}t, 
\end{equation} where $t_0$ is the time in which the excitation of the mode begins. 
Consequently the experimentally obtained shifts are
\begin{equation}
\Delta\lambda_k\propto\alpha_e(\overline{I}_k-\overline{I}_{k-1})=\alpha_e\Delta\overline{I}_k.
\end{equation} 
Our sensor, therefore, can only recognize molecular interactions that keep the analyte molecules confined temporally in the order of $\tau_m$ as well as spatially within the plasmonic hotspots (i. e. near the NR tips).  Molecular processes shorter than $\tau_m$ but occurring repeatedly during $\tau_m$ can be also recognized but with reduced magnitudes, whilst one-time events shorter than $\tau_m$, as for example freely diffusing analyte molecules near the hotspots, are unlikely to be recognized as their $\overline{I}_k$ is significantly lower \cite{baaske2016}.
In order to study the sm-DNA/Pol interactions we take two approaches : The first (Fig. 1c, top) is based on the polymerase interacting with DNA strands immobilized on the NRs (henceforth referred to as the immo-DNA scheme). In this case, the shifts occur due to the changes in $\overline{I}$ as polymerase molecules are attached and detached from the DNA strands, consequently moving in and out of the areas with high field intensities (Fig. 2a, d). For the second approach (Fig. 1c, bottom) the polymerase molecules are immobilized on the NRs (henceforth referred to as the immo-Pol scheme), and the observed shifts are caused by changes in $\overline{I}$ due to the polymerase changing its conformational states accompanied by its interaction with DNA strands (fig. 2b, c, e). Using both approaches we obtain similar transient signal patterns, so-called 'spikes' arising from the sm-DNA/Pol interactions (Fig. 1d). These are comprised of an initial red shift of the resonance position as a molecular interaction starts and a consequent return to the unperturbed mode position (blue-shift) as the interaction ceases. Each individual spike exceeding the wavelength noise $\sigma$ by at least 3 times is found and extracted using a spike detection algorithm \cite{baaske2016}. This algorithm also removes background drifts  as well as returns the average and maximum shifts ($\overline{\Delta\lambda}$, $\Delta\lambda_{max}$) and the spike duration ($\Delta\tau$).
In line with the proofs of the single analyte nature demonstrated in our previous studies \cite{baaske2014single,baaske2016,kim2016}, we have found that the detected spikes from the sm-DNA/Pol interactions  originate from a Poisson process and their detection rates scale linearly with the analyte concentrations (Section 2 of Supplementary Information). We however would like to note that the presented single molecule proofs do not indicate that only one receptor molecule (the immobilized reactant) was monitored overall, as the spikes can originate from an ensemble of receptor molecules immobilized on the NRs. Nonetheless the statistical proof confirms that each individual spike originates from a single receptor interacting with a single analyte molecule, while prior or later spikes may originate from a different receptor.

To further elaborate on the sensor's response we have performed finite element simulations to obtain the near field intensity $I$ of the electric field in the proximity of NRs. For this we used simplified geometry for the polymerase consisting of two moving arms and a stationary bottom with the size parameters obtained via X-ray crystallography \cite{kim2008crystal}. Specifically, we compare the changes in $I$ associated with the movement of the polymerase (fig. 2a, d) as a correspondence to the immo-DNA scheme. For the immo-Pol scheme, the changes of $I$ depending on the variations of angular spread between the thumb and the finger domain of polymerase were compared at two different immobilization positions (fig. 2b, c, e).
\begin{figure}[bth]	
\begin{center}
\includegraphics{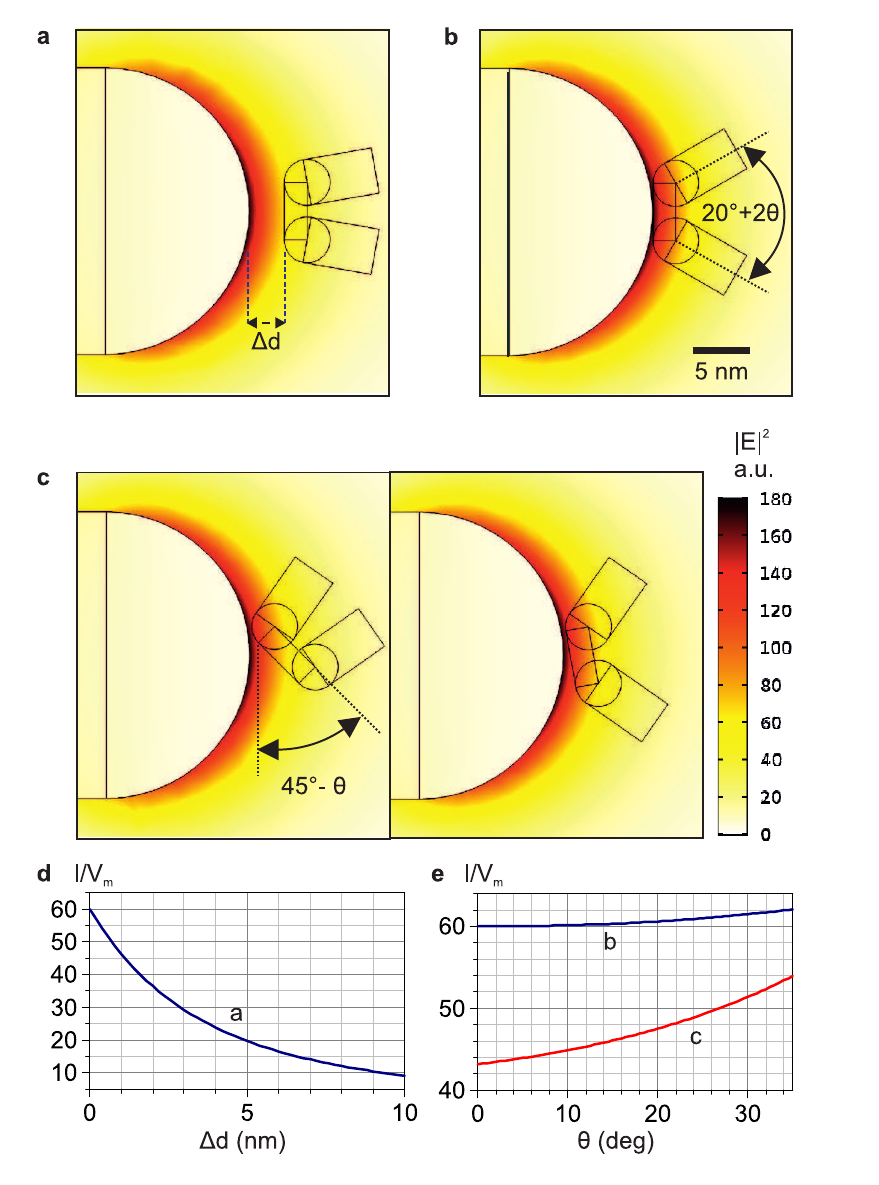}
	\caption{\textbf{Near-field based transduction mechanism.} Panels (a) through (c) show the spatial distributions of the near field's intensity as well as the parameters associated with the respective movement of the polymerases volume: (a) Distance between NR and polymerase. (b) Changing angle between both arms with fixed base. (c) Changing angle between both arms with fixed left arm. (d) and (e) show the dependence of the volume integrated intensity $I$, normalized to $V_m$, on the parameters associated with (a), (b), and (c).}
\end{center}
\end{figure} 
The near-field intensity exhibits a highly inhomogeneous distribution within the volume of the polymerase and rapidly decays with increasing distance from the NR's tip. In consequence $I$ decreases significantly as the gap $\Delta d$ between the NR and polymerase increases on the scale of few nanometers (fig. 2d). Furthermore, $I$ generally increases as the angle $\theta$ between the two arms of polymerase increases while its absolute value as well as $\theta$-dependency largely vary for different immobilization locations (fig. 2e).
Furthermore, we observed that for a different bound position of polymerase an increment of the angle $\theta$ could also lead to a decrease in $I$. This indicates the WGM shift induced by the same conformational change of polymerase (i. e. the same $\theta$) can exhibit a different magnitude as well as a different sign of the shift. In the experiments, however, only the spikes to the red spectral region were observed (Fig. 1d). 
\subsection*{Interaction kinetics of different DNA polymerase species}
\begin{figure*}[htbp]	
\begin{center}
\includegraphics[scale=1]{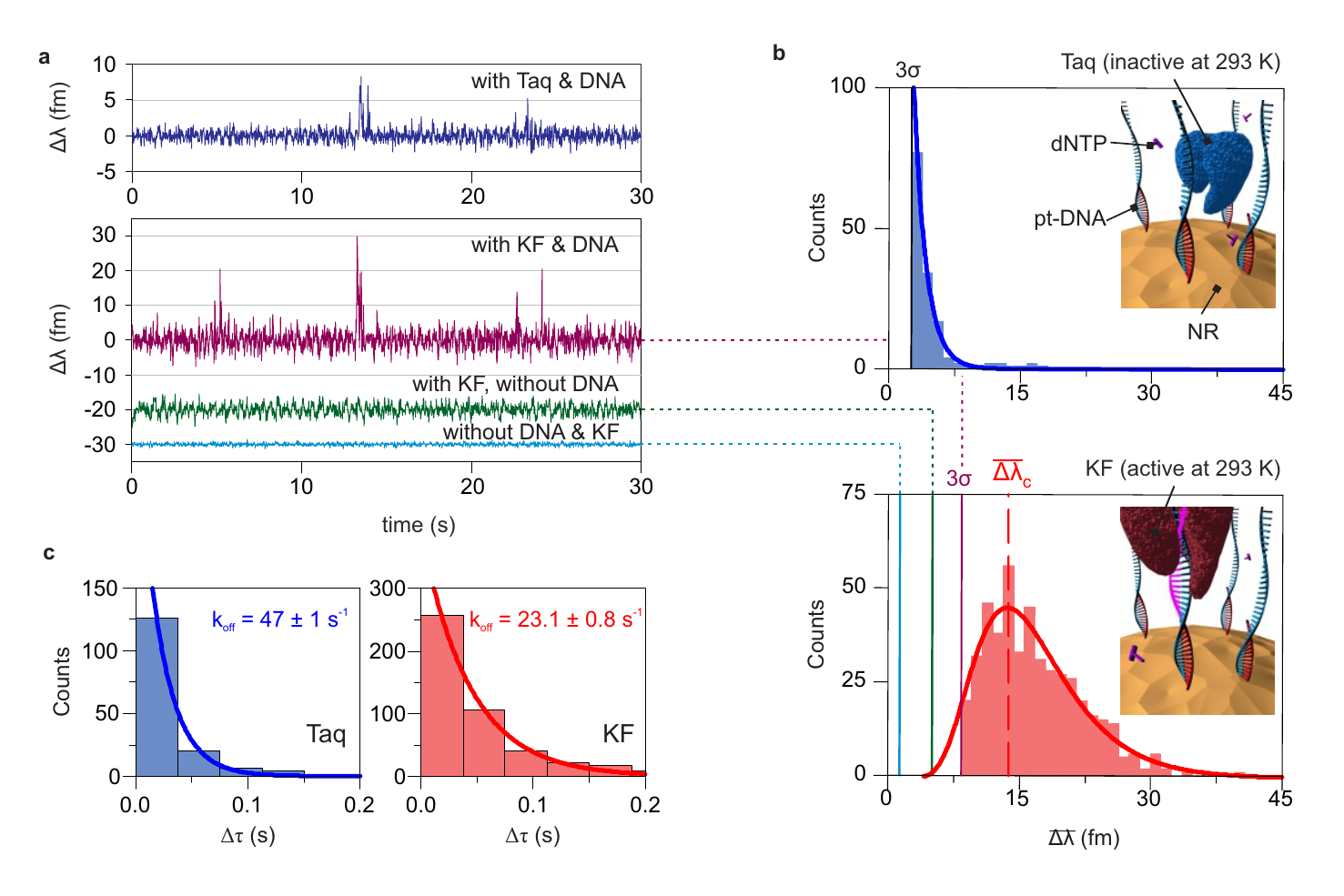}
	\caption{\textbf{sm-DNA/Pol interaction signals using immo-DNA scheme.} (a) Example resonance traces exhibiting spike patterns caused by Taq (top, blue) and KF (bottom, pink) polymerase/DNA interactions and the different noise levels found for pt-DNA functionalized NRs (pink), unfunctionalized NRs (green), and in the absence of KF (light blue). (b) Distributions of the average spike amplitudes $\overline{\Delta \lambda}$ and (c) durations $\Delta \tau$ obtained for Taq (blue) and KF (red) interacting with pt-DNA in the presence of a dNTP concentration of 50 $\mu$M. }
\end{center}
\end{figure*} 
Utilizing the immo-DNA approach (Fig. 1c, top) we experimentally inspect the interaction kinetics of two different polymerase species, Taq and KF at a temperature of $T\approx 293K$. Both species are expected to show apparent differences in their kinetic behavior, as their enzymatic activity is optimal at distinctively different temperatures of $T_{opt}$ = 348 - 353 K and 310 K, respectively. Representative resonance wavelength traces for both species are shown in Fig. 3a. 
In the case of Taq, we do not observe any change in noise level associated with either the presence or the absence of Taq and DNA. As for KF, however, the noise level rises once KF is added and increases even further after DNA is immobilized on the NRs. The former noise increase may be attributed to unspecific short and reversible attachment of KF to the NRs, while the latter might originate from KF interacting with DNA molecules bound to locations on the NRs with low field intensities. Distinct spikes are observed only if DNA is immobilized on the NRs as well as Taq or KF are present in solution, thus confirming the specificity of the monitored sm-DNA/Pol interactions (Section 3 of Supplementary Information).
Furthermore the expected difference in the kinetic behavior of Taq and KF is directly evident by comparing the spike magnitude and duration distributions obtained for both species (Fig 3b, c).
The spike amplitudes found for Taq/DNA interactions exhibit an exponentially decaying distribution (Fig. 3b, top) with 53 $\%$ of the events populating the first bin above the $3\sigma$ limit. In contrast, the spike amplitudes obtained for KF/DNA interactions have a significantly broader distribution exhibiting a clear peak at $\overline{\Delta\lambda}_c$ well in excess of $3\sigma$ (Fig. 3b, bottom).
This stark difference in the distributions is at first surprising, as one would expect higher spike amplitudes for Taq polymerase due to its larger molecular mass (98 kDa) compared to the one for KF (68 kDa). The spike magnitude is however the result of a temporal averaging processes (Eq. 2, 3) and therefore should be seen in conjunction with the spike duration as events shorter than $\tau_m$ are recognized with a reduced shift magnitude. Indeed we find the spike durations $\Delta \tau$ associated with Taq/pt-DNA interactions (Fig 3c, left) to be significantly shorter than those originating from KF/pt-DNA interactions (Fig 3c, right). Correspondingly, both species also yield distinctively different off-rates ($k_{off}$) of $47$\,$\mathrm{s}^{-1}$ and $23$\,$\mathrm{s}^{-1}$ (extracted via fitting of $N(\Delta\tau)\sim e^{-k_{off}\Delta\tau}$ to the respective distributions). The fact that the experiments were performed at 293 K, a temperature rather close to optimal temperature for KF but well below for Taq, indicates there is a correlation between our sensor's signal and the activity of the monitored enzyme. It is however worthwhile to note that the off-rates, extracted from the $\Delta \tau$ distributions, require careful interpretation as they are not necessarily equivalent to the dissociation rates between pt-DNA and polymerase. They rather reflect how long the polymerase resides within the plasmonic hotspots (i. e. a period for which the value $I$ is high enough to be recognized) while it interacts with a DNA strand. This means that $I$ can drop below the recognition threshold before the actual DNA/Pol-interaction has ended, as for instance if a polymerase moves along the over-hanging template strand and away from the NR's surface.
In actuality, our values for $k_{off}$ exceed the dissociation rates reported in other studies \cite{kuchta1987kinetic,christian2009single,langer2015polymerase} by at least one order of magnitude.
It is thus evident that by using the immo-DNA approach the amount of information which is directly obtainable is limited, although it still allowed us to recognize differences in the interaction kinetics of KF associated with ss-DNA and pt-DNA/dNTP interactions (Section 5 of Supplementary Information).
Hence we employ the approach of immobilizing the polymerase on the NRs (the immo-Pol scheme) in the following studies of sm-DNA/Pol interactions.
\subsection*{Conformational transitions of Pfu polymerase at various temperatures}
\begin{figure*}[htbp]	
\begin{center}
\includegraphics[scale=1]{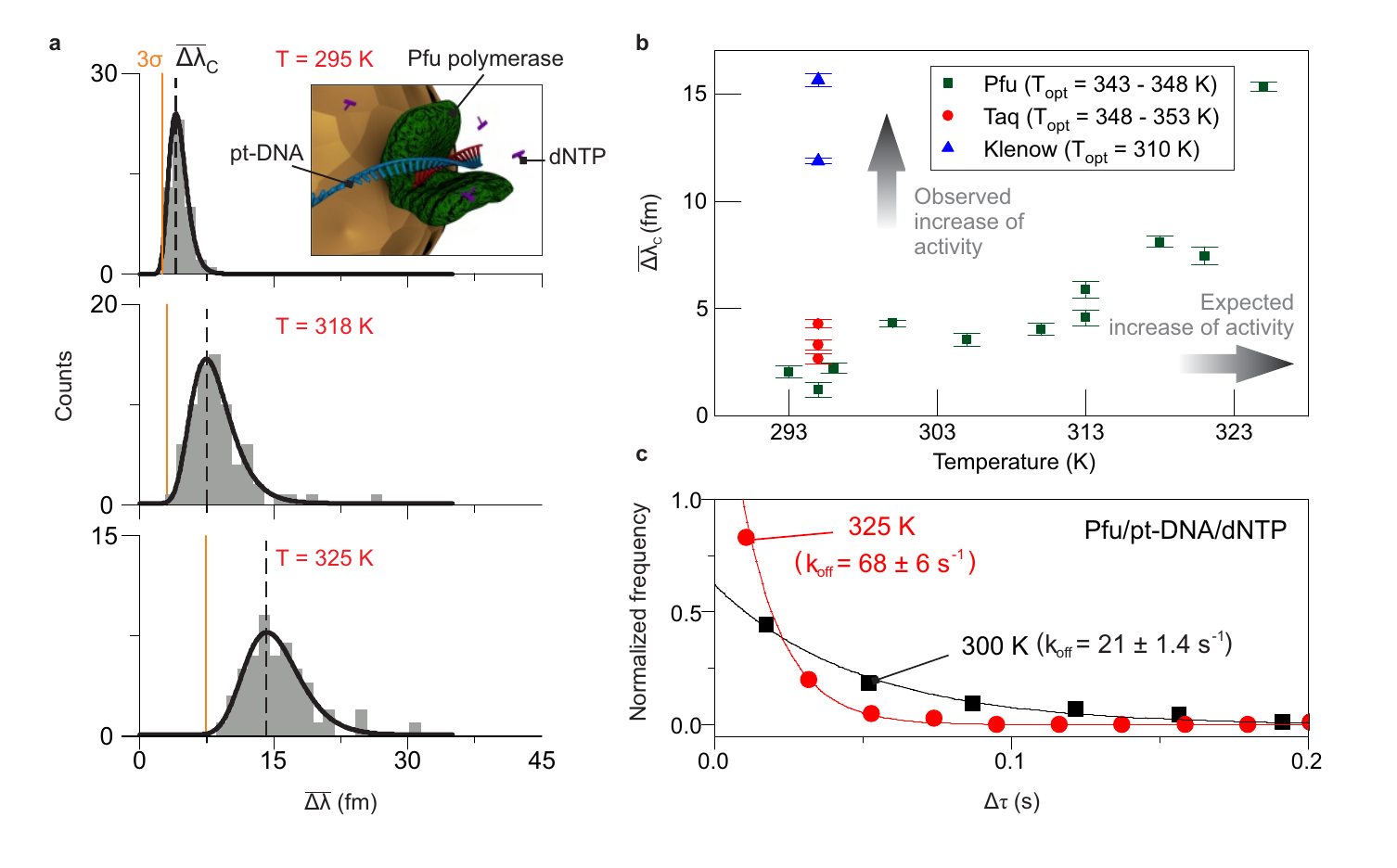}
	\caption{\textbf{sm-DNA/Pol interaction signals using immo-Pol scheme.} (a) Average spike amplitude $\overline{\Delta \lambda}$ distributions obtained for Pfu/pt-DNA/dNTP interactions showing the evolution of overall signal amplitude with increasing temperature and enzyme activity. Peak center positions $\overline{\Delta\lambda}_c$ extracted via log-normal fits (solid lines), indicated by dashed lines. (b) $\overline{\Delta \lambda}_c $ for different DNA polymerase species (Taq, KF, and Pfu) and temperatures. (c) Distributions of spike durations $\Delta \tau$ obtained for Pfu/pt-DNA/dNTP interactions at two different temperatures. The concentration of dNTPs in the solution was kept to 50 $\mu$M.}
\end{center}
\end{figure*}
The above results suggest a possible correlation between our sensor signal, namely the spike amplitudes and durations, and the activity of the monitored enzymes. Nonetheless it is challenging to precisely determine the physical process associated with the signals, as conformational changes and the motion of the whole enzyme cannot be directly separated from the immo-DNA scheme. In order to exclude the polymerase motion's contribution to the signals, we have performed experiments with polymerase immobilized on the NR (immo-Pol scheme as shown in Fig. 2b).
For this we selectively use Pfu polymerase as it maintains its enzymatic activity even when immobilized onto the NR surface. Experimental data confirming its activity, successful immobilization onto the NRs, as well as the fact that the origin of the signals are indeed from the structural change of the Pfu-polymerase are supplied in Section 4, 6, 7 of the Supplementary Information. Furthermore the Pfu-polymerase is thermophilic with its maximum enzymatic activity in the range of $345$ to $348$ K, and hence allows for tuning its activity by adjusting the ambient temperature. In the past studies we had investigated the conformational changes of proteins with respect to the environmental temperature in bulk \cite{kim2015thermal}. This approach is now employed on the single molecule level to study the change in the Pfu/DNA interactions via step-wise increments in the liquid chamber's temperature. The corresponding results for Pfu/pt-DNA/dNTP interactions at different temperatures are showcased in Figure 4. With increasing temperature the center of the peak position $\overline{\Delta\lambda}_c$ shifts toward higher amplitudes and their broadness increases (Fig. 4a). This is line with our previous findings from the comparison of KF- and Taq-polymerase (Fig. 3), indicating that more active polymerases induce larger spike amplitudes (Fig. 4b). As opposed to the results from the comparison of Taq and KF, however, the spike durations observed for the Pfu/pt-DNA/dNTP interactions become shorter as the expected activity of the polymerase increases (i. e. ambient temperature increases) as shown in Fig. 4c. This, together with the increase in the signal amplitudes, signifies that the monitored Pfu polymerase undergoes the pt-DNA/dNTP interaction-induced conformational transitions with increasing magnitudes as well as with increasing speed as the ambient temperature rises. We next compare the temperature dependent Pfu/DNA interactions in the absence and presence of dNTPs (See Fig. 5).  In both cases the center of these peaks shifts toward higher amplitudes with increasing temperature but the dependency is larger if dNTPs are present (Fig. 5a). The overall shift in the center position increases exponentially with temperature as shown in Fig. 4b as well as evident from the corresponding Arrhenius-plots in Fig. 5b, where the dataset obtained in presence of dNTPs exhibits a steeper slope ($-6.5 \pm 0.9$ K) as compared with the one obtained ($-1.7 \pm 0.7$ K) in the absence of dNTPs (numeric values are listed in Table S1 of the Supplementary Information).
\begin{figure*}[htbp]	
\begin{center}
\includegraphics[scale=1]{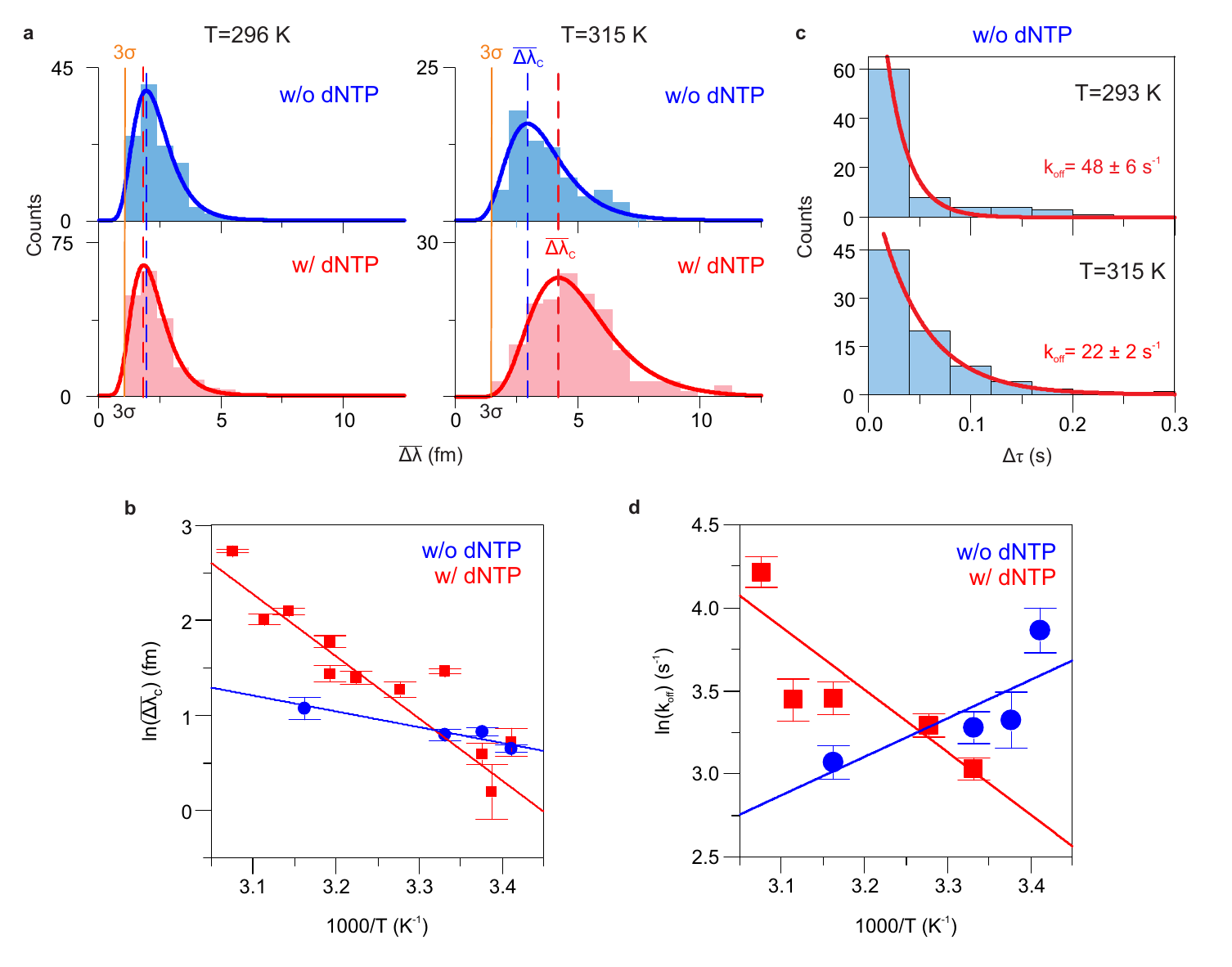}
	\caption{\textbf{Different conformational transitions in the presence and absence of dNTPs} (a) Comparison between average spike amplitude $\overline{\Delta \lambda}$ distributions obtained for Pfu/DNA interactions in the absence (top) and presence (bottom) of dNTPs (of concentration = 50 $\mu$M) at 296 and 315 K. Arrhenius plots displaying the temperature dependence of (b) $\overline{\Delta \lambda}_c $ and off-rates (d) $k_{off}$ found for Pfu in the presence and the absence of dNTPs. (c) Change of the spike duration distributions at two different temperatures in the absence of dNTPs. }
\end{center}
\end{figure*}
The temperature dependence of the spike durations, in contrast to what was observed for the spike amplitudes, displays a distinctive difference with respect to the presence of dNTPs: In the presence of dNTPs the spike durations decrease with increasing temperature as shown in Fig. 4c, while in the absence of dNTPs the spike durations increase with increasing temperature (Fig. 5c). In both cases the $k_{off}$ values extracted from the duration distributions show an exponential dependence on temperature, although with opposite directions as is evident from the  Arrhenius plot in Fig. 5d with the corresponding slope values of $-3.8 \pm 0.6$ and $2\pm 1$ K for with and without dNTPs.

In conjunction we can conclude that in the absence of dNTPs, the Pfu polymerase undergoes a conformational transition, triggered by its interaction with pt-DNA, such that $\Delta\overline{I}$ increases with increasing temperature. This can be caused either by a larger fraction of the enzyme's volume contributing to the movement or a constant fraction moving over a greater distance. Furthermore without dNTPs the enzyme remains longer in the observed conformation as the ambient temperature increases. This could be interpreted as the polymerase transitioning to a state in which it waits for dNTPs to arrive, where the energy necessary to maintain the state is thermally provided.
In the presence of dNTPs, the temperature dependence of the spike amplitudes follows the same behavior and hence can be explained similarly with respect to $\Delta\overline{I}$. The comparably larger temperature dependence of the shifts found in the presence of dNTPs, therefore, could be related to the 
the conformational change (which is triggered by the pt-DNA) involving
additional movements (either they originate from the movement from an additional fraction of volume or a larger degree of movement) induced by dNTPs. The reduction in spike duration with increasing temperature could be understood such that the additional process triggered by dNTPs accelerates and the overall Pfu/DNA interactions is consequently shortened as more thermal energy is supplied.

These results, especially the distinctively different temporal behavior in the presence and the absence of dNTPs, provide strong evidence that the monitored conformational changes, accompanied by the Pfu/DNA-interactions, involved the incorporation of nucleotides. They however do not imply individual spikes are associated with only one incorporation process. Given the fact that the duration of our spikes is also not in the range which would be expected for transcription of the whole template strand (in the order of a few second), it is more likely that we observe an initial conformational change as the Pfu polymerase forms the complex with the pt-DNA accompanied by possibly few nucleotide incorporations. 

\subsection*{Discussion}
We have demonstrated our sensor's capability to detect polymerase/DNA interactions linked to enzymatic activity on a single molecule level by 
monitoring the kinetics as well as conformational transitions of DNA polymerases. Our results exhibit a clear correlation between our sensor's signal and the enzymatic activity of three different types of polymerase at various ambient temperatures. Moreover we have shown that the magnitude and the duration of the conformational changes of polymerase associated with DNA interactions vary with respect to the presence and the absence of dNTPs. In this context we have found a distinct difference in the temperature dependence of the enzymes' kinetic behavior, providing the strong evidence for the correlation of our sensor's signal with enzymatic activity in the form of nucleotide incorporation. 
Our results are potentially significant as monitoring enzymatic activity in a multiplexed and high throughput fashion is a crucial requirement for next-generation sequencing. The fact that our approach is label-free and the sensor signals can be monitored in real time opens a new and direct way to determine conformational substates of various proteins without the necessity to mitigate label associated background fluctuations. Our sensor, however, is intrinsically limited by our signal amplification method as the plasmon enhanced near field may only probe a fraction of the enzymes volume. This limitation might yet become an advantage as it may, in turn, allow for the selective probing of certain protein subdomains. Furthermore, enzymatic kinetics can be easily tested with respect to diverse medium conditions such as molecule concentrations, ionic strength, pH, and temperature. In this context, combining our method with label-based techniques would be promising when linked to the extensive knowledge already established via label-based methods, therefore diversifying quantitative analysis. For example, in combination with FRET, our sensor would allow for probing of conformational changes beyond FRET's spatial limitations yet take advantage of its selectivity. 
\subsection*{Methods}
All solutions without NRs were filtered with $0.1$ $\mu$m syringe filters (Merck Millipore) prior to usage.
\subsubsection*{The immo-DNA recognition scheme}
Sensor assembly is conducted by adopting a modified version of the three-step wet-chemical procedure used in previous works \cite{baaske2014single}.
Step (1): cethyltrimethylammonium bromide (CTAB)-stabilized gold NRs with diameters of $10\,\mathrm{nm}$ and lengths of $35\,\mathrm{nm}$ (Nanopartz) are immobilized on the microresonator surface. For this the NRs are injected into a sample cell holding about $0.5$ mL of $100\,\mathrm{mM}$ NaCl solution at pH $\approx 1.6$.
This process is directly monitored and individual NR binding events are classified as discrete steps in the resonance position and linewidth traces. 
Step (2): thiol-modified single-stranded DNA ([ThiC6]-5'-TTTTCTCGTTGGGGTCTTTGCTC, Eurofins) are conjugated to the adsorbed NRs.
In order to cleave any disulfide bond the thiolated DNA is treated with $100\,\mathrm{mM}$ dithiolthreitol (DTT) and $100\,\mathrm{mM}$ NaCl for $30\,\mathrm{min}$ at room temperature \cite{nicewarner2002hybridization} prior to measurement.
The NR-modified sphere is then immersed in a solution with $500\,\mathrm{mM}$ NaCl, $0.02\,\%$ (w/w) sodium dodecyl sulfate (SDS) at pH $\approx 3$ and $1\,\mu\mathrm{M}$ DNA \cite{baaske2014single,shi2013facile}. The time spans required to produce a surface coverage that is sufficient for monitoring sm-DNA/Pol interactions are in the range of 10 to 30 minutes, albeit longer reaction times may result in undesirably high DNA surface densities and hinder protein-DNA interactions. 
Step (3): In NEBuffer 2 (New England Biolabs Inc.), the initial pH of the solution is reduced to $6.7$ by adding $2\,\mathrm{mM}$ HCl as to maintain stable NR adsorption. 
If required, ss-DNA was hybridized to pt-DNA by injection of the template strand DNA (60-nt, CCGACAACCACTACACCGGTCTGAGCACCCAGTCCGCCCTGAGCAAAGACCCCAACGAGA), then followed by the introduction of the desired amount of polymerase (i.e. Taq and KF, New England Biolabs Inc.).
\subsubsection*{The immo-Pol recognition scheme}
Sensor assembly is achieved according to the following steps.
Step (1): polymerase-gold nanorods conjugates were prepared by mixing 2 $\mu$L \textit{Pfu} DNA polymerase (BioVision) from a 2.5 units$/\mu$L stock solution with 5 $\mu$L of a solution containing citrate gold nanorods (25 nm diameter and 49 nm length, Nanopartz) at a concentration of $5.7\cdot 10^{11}$ nanoparticles/$\mu$L.
Step (2): The surface of the freshly fabricated microsphere is functionalized with aminopropyltriethoxysilane (APTES), (C$_9$H$_2$3NO$_3$Si)  by immersing the microsphere in a 100 $\mu$L droplet of 2.5 (v/v) \% APTES for about 1-2 minutes. The binding of Pfu-NR conjugates to the microsphere is then performed in a sample chamber filled with PCR buffer. The PCR buffer contains 10 mM Tris-HCl, 50 mM KCl, 1.5 mM MgCl$_2$, and 0.001 $\%$ (w/v) Gelatin (Sigma Aldrich) . Step (3): pt-DNA used for the measurements was prepared by mixing the template and primer strands with a 1:1 molarity ratio in a solution containing 50 mM NaCl. This mixture is heated to 94 $^{\circ}$C with an Eppendorf incubator and cooled down to room temperature before use. The observation of pt-DNA and immobilized Pfu polymerase was then undertaken in the PCR buffer.
\subsubsection*{Numerical analysis}
Numerical simulations were performed using COMSOL Multiphysics (frequency domain module). The polymerase was modeled as three rectangular parallelepipeds (6 x 4 x 5 nm for the moving arms and 2 x 5 x 5 nm for the stationary bottom) which are conjugated with two cylinders (both 2 x 5 nm) to maintain a constant volume while increasing the angle between two arms. It was initially attached to the center of the gold NR's tip and this gold NR was modeled as a prolate circular cylinder with hemispherical end caps with dimensions of 25 x 49 nm. The refractive index of polymerase and the surrounding medium (water) used for the simulation were 1.667 and 1.332, and the frequency-dependent refractive index values of gold were taken from \cite{johnson1972optical}. The simulation domain was surrounded by a perfectly matched layer (PML) to absorb the outward propagating radiation. To calculate the local field enhancement, the entire domain was illuminated with a plane wave at a pump wavelength of 642 nm. The polarization state of the incoming field was parallel to the long axis of the NR. With regards to meshing, a swept mesh was used to discretise the external and PML domain. Note that NR as well as polymerase domains were meshed using free tetrahedral discretisation, while finer meshes were used in the NR and polymerase domains. 
\subsection*{Acknowledgments}
The authors acknowledge financial support for this work from the Max Planck Society and give thanks to prof. Dr. Christian Koch from the Biochemistry chair of Friedrich-Alexander University Erlangen-N\"urnberg for providing access to the PCR instrument. E.K. and M.D.B. thank S. Vincent for his feedback on the manuscript.
\subsection*{Author contributions}
E.K., M.D.B., I.S., and P.S.W. contributed equally. E.K. and M.D.B. wrote the manuscript and performed the simulation. M.D.B. developed the experimental setup as well as data analysis tools and supervised the experiment. P.S.W, I.S., and E.K. performed the experiment. F.V. supervised the entire project. All authors commented on the manuscript.
\subsection*{Additional information}
The authors declare no competing financial interests. 
\bibliographystyle{naturemag}
\bibliography{ArxivSubmission}
\end{document}